\documentstyle[epsfig]{aipproc}

\begin{document}

\title{Far-Infrared and Submillimeter Observations of High Redshift Galaxies}
 
\author{David A. Neufeld}
\address{Department of Physics and Astronomy, The Johns Hopkins University,
3400 North Charles Street, Baltimore, MD 21218}

%\lefthead{LEFT head}
%\righthead{RIGHT head}
\maketitle

\begin{abstract}

Observations at far-infrared and submillimeter wavelengths promise
to revolutionize the study of high redshift galaxies and AGN by 
providing a unique probe of the conditions within heavily extinguished 
regions of star formation and nuclear activity.  Observational 
capabilities in this spectral region will expand greatly in the
next decade as new observatories are developed both in space and on the 
ground.  These facilities include the Space Infrared Telescope
Facility  (SIRTF), the far-infrared and submillimeter telescope (FIRST)
and the millimeter array (MMA).  In the longer term, the requirements 
of high angular resolution (comparable to that of HST), full wavelength
coverage, and high sensitivity 
(approaching the fundamental limit imposed by photon counting statistics) 
will motivate the development of far-IR and submillimeter 
space interferometry using cold telescopes and incoherent detector
arrays.

\end{abstract}

\section*{Introduction}

Key scientific questions about the Universe that have been raised at 
this meeting and elsewhere include

\vskip 0.1 true in

$\bullet$ What is the history of star formation in the Universe?

\vskip 0.1 true in
$\bullet$ What is the history of metallicity and dust content in the
Universe?

\vskip 0.1 true in
$\bullet$ What is the origin of the extragalactic background observed by
the DIRBE experiment on COBE?

\vskip 0.1 true in
$\bullet$ What are the relative contributions of stars and of active galactic
nuclei to the luminosity of the Universe, and how do they vary with redshift?

\vskip 0.1 true in

In this paper, I will argue that observations in the far-infrared and
submillimeter wavelength region (40 -- 1000 $\mu$m, corresponding to
rest wavelengths in the range 5 -- 300  $\mu$m for galaxies at $z = 2 - 5$) 
offer a unique probe of the high redshift Universe that will address 
these questions.  The happy coincidence of several astronomical facts 
make far-infrared and  submillimeter observations particularly powerful.  

{\bf First, galaxies are extremely luminous at far-infrared rest wavelengths.} 
The spectrum of our own Milky Way galaxy, for example, shown in Figure 1, 
exhibits two distinct peaks, the first at around 1 $\mu$m resulting from 
the integrated emission from stars, and the second at around 100 $\mu$m 
resulting from interstellar dust emission.  The representation given here 
(in which equal areas correspond to equal rates of photon emission)
shows that most of the Galaxy's {\it photons} emerge in the far-infrared region.
The Milky Way is entirely unremarkable in this regard: indeed, in many
starburst galaxies even the {\it energy} output is dominated
by far-infrared radiation.  The strength of the far-infrared emission from 
galaxies simply reflects that fact that the average visual extinction is 
sufficient to allow a significant fraction of the starlight to be reprocessed
by interstellar dust.
\begin{figure}
\centerline{\epsfig{file=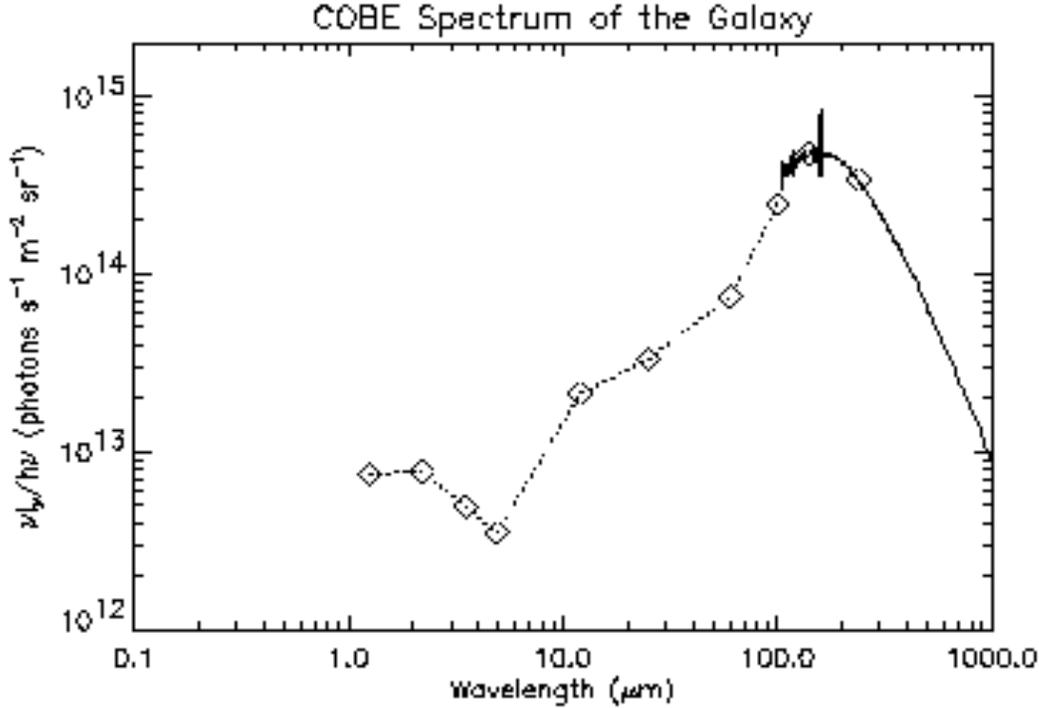,width=6.0in,height=4.0in}}
\vspace{10pt}
\caption{COBE spectrum of the Milky Way (from Mather et al.\ 1998)}
\end{figure}

{\bf Second, the opacity of interstellar dust is a strongly decreasing function
of wavelength}, allowing embedded regions of star formation and nuclear
activity that are invisible at optical and even near-infrared wavelengths to
be detected in the far-infrared and submillimeter spectral regions.
A second implication of the strong wavelength-dependence of the dust opacity 
is that the submillimeter region provides a unique cosmological window to the
high redshift Universe, the background caused by dust emission
from $z \sim 0$ (local galaxies) dropping 
rapidly with increasing wavelength longward of $\sim 200\,\mu$m and the
background from $z=1500$ (the CMB) dropping rapidly with decreasing 
wavelength shortward of $\sim 800\,\mu$m.

{\bf Third, the 5 -- 300 $\mu$m (rest) wavelength range is extremely 
rich in atomic and molecular diagnostics} that can serve as powerful probes 
of the physics and chemistry of interstellar gas and dust.  The remarkable
richness of the mid- and far-infrared spectral region is demonstrated
by the Infrared Space Observatory (ISO) spectrum of the Orion region (van
Dishoeck et al.\ 1998) in Figure 2, which shows numerous rotational lines
of H$_2$ and H$_2$O, fine structure emissions from a wide variety of
atomic ions, as well as several broader features associated with
interstellar dust.  In the next section, I will discuss the potential
importance of such spectral features in constraining the properties of
high-redshift galaxies.

\begin{figure}
\centerline{\epsfig{file=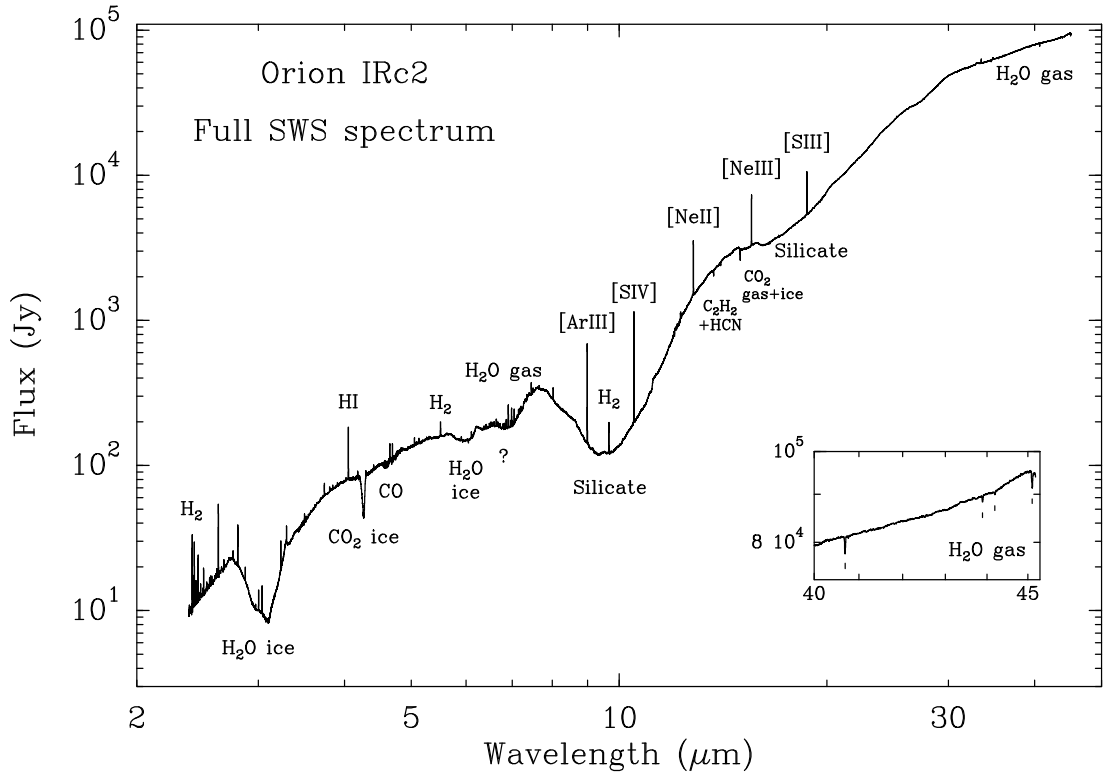,width=4.7in,height=6.2in,angle=270}}
\vspace{10pt}
\caption{
ISO Short Wavelength Spectrometer spectrum of Orion IRc2,
from the paper of van Dishoeck et al.\ (1998).}
\end{figure}

\section*{Emission mechanisms at far-infrared and submillimeter wavelengths}

{\bf Interstellar dust} is the dominant source of far-infrared and submillimeter 
radiation in galaxies.  Recent SCUBA observations of dust continuum radiation 
-- reported, for example,  at this meeting by Ian Smail
-- have already demonstrated
the power of submillimeter observations to probe the Universe at high redshift.
Some -- although by no means all -- of the identified SCUBA sources are 
galaxies at redshifts $z > 2$ (e.g.\ Ivison et al.\ 1998), and the 
$\sim 25\%$ of SCUBA sources for which 
no optical counterpart can be found may well be sources at very
high redshift  (or alternatively low-redshift galaxies or AGN 
that are very heavily 
extinguished).   
Although the field is currently in its infancy, observations of dust continuum 
radiation will ultimately allow the the effects of dust absorption to be
corrected for quantitatively in models for the luminosity history of the
Universe, and will elucidate the relative contribution of sources at 
different redshifts to the  extragalactic background detected by the DIRBE 
experiment on COBE (Hauser et al.\ 1998).

Although it shows a continuum spectrum, emission from dust is not featureless.  
It exhibits several broad features of large equivalent width, most notably 
the silicate feature at 9.7 $\mu$m and several bands in the 3.3 -- 11.3 
$\mu$m range that have been attributed to polycyclic aromatic hydrocarbons
(Allamandola, Tielens \& Barker 1985), 
and these features may allow redshifts to be estimated from far-infrared 
observations of very modest spectral resolving power.  

{\bf Interstellar gas} emits a rich spectrum of atomic and molecular
line radiation in the  5 -- 300 $\mu$m range, which -- although a negligible 
contribution to the overall far-infrared and submillimeter emission -- 
dominates the cooling of the interstellar gas and provides valuable 
diagnostics of the physical and chemical conditions.  

Fine-structure emissions from the low-ionization species C$^+$
and O dominate the cooling of neutral atomic gas clouds.  The C$^+$ 
$^2P_{3/2} - ^2P_{1/2}$ line at 158~$\mu$m has an upper state energy
($E_u/k$) corresponding to only 92~K and is therefore readily excited
in cold atomic clouds.  In most galaxies, the C$^+$ 158~$\mu$m line
is the strongest source of line emission and accounts for 0.2 -- 1$\,\%$ 
of the total far-infrared luminosity (Malhotra et al.\ 1997), 
this percentage representing
the fraction of the absorbed radiant energy from stars
that is deposited in the interstellar gas rather than the dust.\footnote{Note, 
however, that the 
{\it relative} strength of the C$^+$ 158~$\mu$m line is considerably smaller
in those galaxies that show the strongest far-infrared continuum emission.
Malhotra et al.\ (1997) have argued that this effect likely arises because 
the larger ultraviolet fluxes incident upon cold clouds within such galaxies 
lead to larger positive charges on the interstellar dust grains and a
resultant decrease in the efficiency of grain photoelectric emission that is the
primary mechanism for heating the gas.}
Spectroscopic observations of the C$^+$ 158~$\mu$m line along with the
O 63$\mu$m and 145$\mu$m lines ($^3P_1 - ^3P_2$ and $^3P_0 - ^3P_1$ with
$E_u/k = 227$ and 326 K respectively) from high-redshift galaxies will
yield reliable redshifts and will allow the heating rate for the
interstellar gas to be determined.  

In dense regions of the interstellar medium that are well
shielded from starlight, the gas is primarily molecular and its emission is
dominated by rotational emissions from molecules.  Gas temperatures
in molecular regions range from $\sim 10\,\rm K$ in quiescent clouds to
several hundred Kelvin in gas that has been heated by a nearby 
star or protostar, or even several thousand Kelvin in shocked regions.
The radiative cooling of molecular clouds is an essential feature of the star 
formation process, because cloud collapse involves the conversion of gravitational
potential energy to thermal energy and can proceed only if the latter
is efficiently removed.   Theoretical calculations (e.g.\ Neufeld, 
Lepp \& Melnick 1995) predict that over a wide range of physical conditions
the radiative cooling of molecular gas is dominated by rotational 
transitions of the molecules H$_2$, CO and H$_2$O in the 
7 -- 600 $\mu$m region.   At low temperatures, submillimeter
transitions of CO are the primary coolant, while at higher temperatures
pure rotational lines of H$_2$ (e.g.\ the S(0), S(1), S(2), S(3), S(4) and S(5)
lines at 28.3, 17.0, 12.3, 9.66, 8.03, 6.91 $\mu$m)
and of H$_2$O (many lines in the 40 - 600 $\mu$m region)
are expected to dominate the cooling.
This prediction is corroborated by recent
ISO observations of H$_2$ and H$_2$O emissions from nearby regions
of star formation (e.g.\ van Dishoeck et al.\ 1998, see Figure 2; 
Harwit et al.\ 1998)
and well as by extensive ground-based observations of CO carried out
previously toward both nearby and high-redshift galaxies (e.g. Omont et al.\ 1996).  
Measurements of line ratios permit the density, temperature, and molecular abundances 
within the molecular gas to be constrained.

In addition to probing cold atomic and molecular gas,
far-infrared and submillimeter observations of high redshift galaxies
also promise to yield invaluable information about photoionized
regions.  Many galaxies are luminous sources of mid IR fine 
structure emissions from NeII (12.8 $\mu$m), OIII (52, 88 $\mu$m), 
NeIII (15.6, 36.0 $\mu$m), NeV (14.3, 24.2 $\mu$m) and several
other ions that result from photoionization by radiation shortward 
of the Lyman limit.  Such mid-IR lines provide unique information about 
the metallicity and gas density in ionized regions, as well 
the spectral shape of the ionizing radiation field (e.g.\ Voit 
1992).  These transitions show several 
important advantages over the optical wavelength lines 
traditionally used to study HII regions: they are not heavily 
extinguished by interstellar dust; their luminosities are
only weakly dependent on temperature and therefore provide
model-independent estimates of metallicity; and they provide
line ratios that are useful diagnostics of density over
a wide dynamic range (e.g. Spinoglio \& Malkan 1992).  The
availability of rare gas elements (e.g. Ne, Ar) allows the 
metallicity to be determined without the complicating effects of 
interstellar depletion, and the availability of a wide range of 
ionization states (e.g. NeII and NeV) provides an excellent 
discriminant between regions that are ionized by hot stars and those that are
ionized by a harder source of radiation such as an AGN.  The power of
that discriminant has been demonstrated by the recent ISO observations
shown in Figure 3 (Moorwood et al.\ 1996).  Here the otherwise
similar spectra of the starburst galaxy M82 and the Circinus galaxy
(which contains an active nucleus) are distinguished by the
presence in Circinus of a variety of highly ionized species
such as NeV and Ne VI that can only result from a very hard 
source of ionizing radiation.

\begin{figure}
\centerline{\epsfig{file=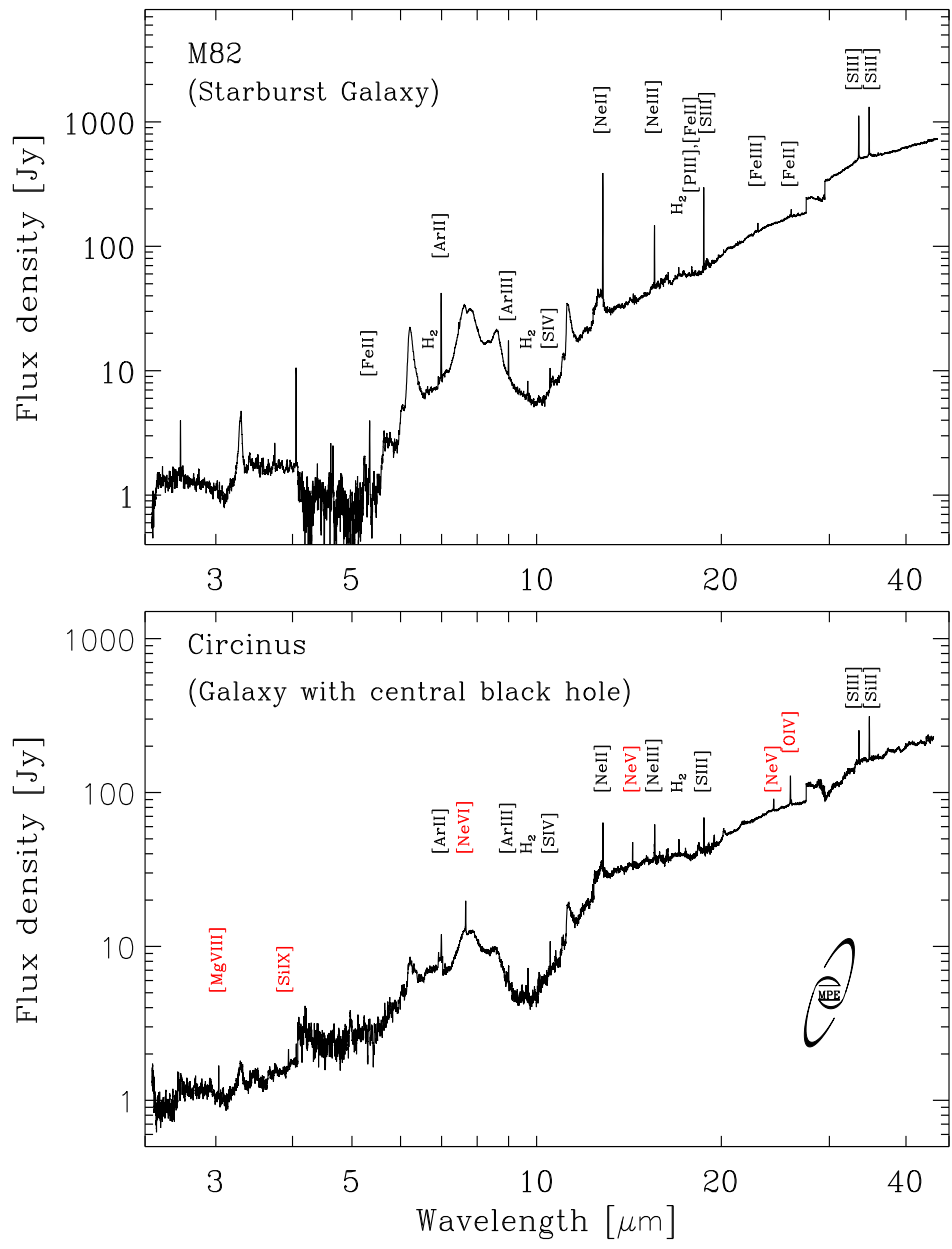,width=5.0in,height=7.0in}}
\vspace{10pt}
\caption{
Comparison of ISO Short Wavelength Spectrometer
spectra of M82 and Circinus, demonstrating the 
power of mid-IR fine structure lines as discriminants
of the ionizing spectrum 
(from the ISO science gallery, credit: ESA/ISO, SWS, Moorwood; 
see also Moorwood et al.\ 1996)}
\end{figure}

Figure 4, from the review paper of Hollenbach \& Tielens (1997),
is a schematic representation of the interaction between starlight and
the interstellar medium that summarizes the various far-infrared and 
submillimeter emission mechanisms described above.  Most of the radiant
energy from starlight is deposited at moderate visual extinctions, 
$A_V \sim 1$.  Roughly 99$\%$ of the starlight heats the 
interstellar dust and is reprocessed as far-infrared continuum 
radiation, while very roughly $1\%$ heats the gas and is reprocessed as line 
radiation (primarily the C$^+$ 158 $\mu$m line).  At visual extinctions 
$A_V > 3$ (rightmost region), the gas is primarily molecular, and the gas 
cooling is dominated by infrared and submillimeter rotational lines of 
molecules,
particularly H$_2$, CO, and H$_2$O.  In unshielded regions of high ultraviolet
flux (leftmost region), the gas is highly ionized, and the cooling is dominated
by optical and ultraviolet line emission.  In this zone, mid-infrared 
fine structure lines, although not the major coolant, are powerful
diagnostic probes.

\begin{figure}
\centerline{\epsfig{file=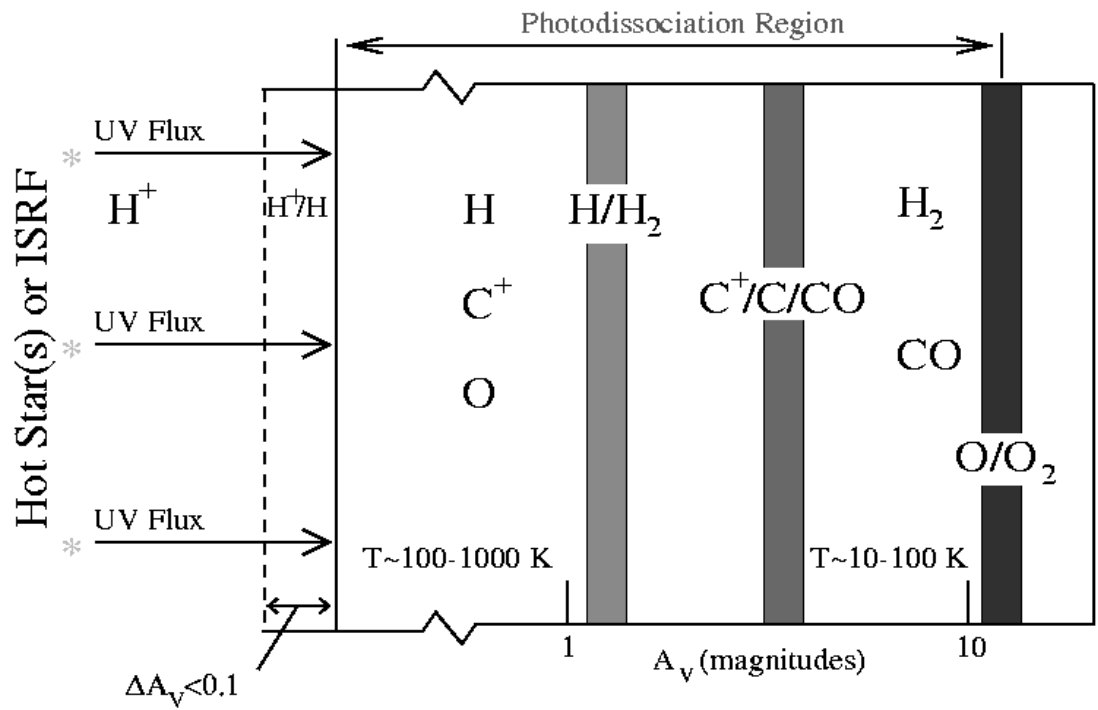,width=6.0in,height=4.0in}}
\vspace{10pt}
\caption{
Interaction of starlight with the interstellar medium,
from the review of Hollenbach \& Tielens (1997).}
\end{figure}

\section*{Observational capabilities in the next decade}

The next decade (2001 -- 2010) promises substantial improvements in
observational capabilities in the far-infrared and submillimeter spectral
regions, thanks to several new observatories that are expected to begin 
operations.  My goal in this article is not to give a comprehensive
review of all these new facilities but rather to discuss briefly selected 
observatories that will offer capabilities most directly relevant to the 
study of galaxies at high redshift.

The {\bf Space Infrared Telescope Facility} (SIRTF)\footnote{The SIRTF
home page is at http://sirtf.jpl.nasa.gov}, 
scheduled for launch 
at the end of 2001, will deploy a liquid helium cooled
85 cm telescope capable of carrying
out observations of extremely high sensitivity.  The Multiband
Imaging Photometer (MIPS) instrument on
board SIRTF will offer diffraction-limited imaging using sensitive detector
arrays at wavelengths of 24, 70 and 160 $\mu$m, as well as very low 
resolution ($\lambda / \Delta \lambda \sim 10$) spectroscopy in the 50 -- 100 
$\mu$m region.  The principal limitation of SIRTF for the detection of
galaxies at far-infrared wavelengths is the large size of the diffraction-limited
beam:  particularly at 160 $\mu$m, MIPS will be source confusion
limited for observations of relatively short duration.  The Infrared Spectrograph (IRS) 
instrument will be capable of moderate resolution spectroscopy ($\lambda / \Delta \lambda 
\sim 600$) over the 10 -- 37 $\mu$m range, with a large spectral multiplex
advantage that will allow full, high-quality spectra to be obtained very much 
more quickly than was possible with ISO.  While the wavelength coverage
of IRS does not quite reach the 40 -- 1000 $\mu$m range that is the subject of this
article, IRS deserves mention here because of its capability for detecting
mid-infrared line emission from ions in HII regions.

The {\bf Far Infrared and Submillimeter Telescope} (FIRST)\footnote{The FIRST
home page is at http://astro.estec.esa.nl/SA-general/Projects/First/first.html}
will be a space observatory with a much larger ($\sim 350$~cm) primary mirror, but one 
that is not actively cooled.   Current plans call for the launch of FIRST 
in 2007, with instrumentation capable of carrying out broad band photometry,
imaging spectroscopy, and high-resolution heterodyne spectroscopy.  The 
wavelength coverage will extend to much longer wavelengths than SIRTF,
allowing a far wider range of atomic and molecular line emissions to be
studied spectroscopically.   Again, the relatively large diffraction limit
at these wavelengths for any single dish instrument of reasonable size means
that source confusion will be significant except for observations of short
duration (e.g.\ Blain, Ivison \& Smail 1998).
Thus interferometers will be critical for the study
of all but the most luminous galaxies at high redshift, and the most important 
impact of SIRTF and FIRST on studies of high redshift galaxies is likely 
to be in measuring spectra of low redshift galaxies that can be used as 
templates for understanding future interferometric observations.

The {\bf Millimeter Array} (MMA)\footnote{The MMA
home page is at http://www.mma.nrao.edu}
will have an extremely powerful 
interferometric capability, providing spatial resolution as good as 
$\sim 0.01 ^{\prime \prime}$, wavelength coverage down
to 350 $\mu$m, and high spectral resolution.  MMA promises
to allow large 
numbers of high redshift sources to be detected routinely and associated 
unambiguously with  optical counterparts.  It will make use of $\sim 36$ antennae of
diameter $\sim 10$~m that can be deployed over baselines of several
kilometers on a high plateau site in Chile.  An observatory of more modest
collecting area -- the Smithsonian Astrophysical Observatory's Submillimeter
Array (SMA) -- will operate in a Northern Hemisphere site (Mauna Kea).
The primary limitations of these ground-based facilities are those imposed 
by Earth's atmosphere, which only permits observations in a series of submillimeter
windows all longward of 300$\,\mu$m, and by
the fundamental sensitivity limits set by heterodyne receivers and warm telescopes.

\section*{The longer term: far-infrared and submillimeter interferometry 
from space}

The ideal instrument for the study of far-infrared and submillimeter 
emissions from high redshift galaxies would combine (1) full wavelength
coverage; (2) HST-like spatial resolution; (3) sensitivity approaching
the fundamental limit imposed by photon-counting statistics; (4) high spectral 
resolution ($\lambda / \Delta \lambda $ of at least $10^4$).  The
first and third of these capabilities require a space observatory; the second
requires interferometry; and the third requires a cooled telescope
(barely warmer than the CMB) equipped with a new (not presently
existing) generation of incoherent detectors rather than heterodyne receivers; 
and the fourth  can be accomplished by means of a Fabry-Perot or Michelson 
interferometer.  

In a recent white paper (Mather et al.\ 1998), we have presented a
preliminary study of such an instrument -- dubbed the Submillimeter
Probe of the Evolution of Cosmic Structure, SPECS\footnote{The SPECS
home page is at http://www.gsfc.nasa.gov/astro/specs} -- in which we
envisaged a Michelson interferometer providing
spatial and spectral interferometry with three, cold, free-flying elements 
of diameter $\sim 3$~m deployable over baselines $\sim 1\,\rm km$.  
Although such a facility may lie significantly beyond what could be built today, Mather 
et al.\ 1998 have emphasized  the importance of developing key technologies over 
the next decade to make such an instrument feasible in the decade 2011 -- 2020; 
those technologies include formation flying, active cooling of large mirrors,
and the development of sensitive incoherent detector arrays.  In particular,
photon-counting incoherent detectors  -- which do not yet
exist at these wavelengths but would likely be some
type of superconductive device -- would offer enormous sensitivity advantages for 
faint sources both over current bolometers and relative to the fundamental limit of 
a heterodyne receiver.\footnote{A perfect
photon-counting detector is more sensitive than a perfect heterodyne receiver 
by a factor $\sim (\Delta \nu /R)^{1/2}$, where 
$\Delta \nu$ is the bandwidth and $R$ is the photon arrival rate, a factor 
much larger than unity for faint extragalactic sources.}

I gratefully acknowledge the support of a grant from NASA's Long Term
Space Astrophysics Research Program.  I thank Ewine van Dishoeck
and David Hollenbach for making available Figures 2 and 4.
It is a pleasure also to acknowledge
helpful discussions with Mark Voit, Harvey Moseley and John Mather.


\begin{references}


\bibitem{}
Allamandola. L., Tielens, A.G.G.M., \& Barker, J.R. 1985, ApJ, 290, L25

\bibitem{}
Blain, A.W., Ivison, R.J., \& Smail, I. 1998, MNRAS, 296, L29

\bibitem{}
Harwit, M., Neufeld, D.A., Melnick, G.J., \& Kaufman, M. 1998, 
ApJ, 497, L105

\bibitem{}
Hauser, M.G., et al. 1998, ApJ, 508, 25

\bibitem{}
Hollenbach, D.J., \& Tielens, A.G.G.M. 1997, ARA\&A, 35, 179

\bibitem{}
Ivison, R.J., Smail, I., Le Borgne, J.-F., Blain, A.W., Kneib, J.-P., 
Bezecourt, J., Kerr, T.H., \& Davies, J.K. 1998, MNRAS, 298, 583

\bibitem{}
Malhotra, S., et al. 1997, ApJ, 491, L27

\bibitem{}
Mather, J.C., et al.\ 1998, astro-ph/9812454

\bibitem{}
Moorwood, A.F.M., Lutz, D., Oliva, E., Marconi, A.,
Netzer, H., Genzel, R., Sturm, E. \& de Graauw, T. 1996, 
A\&A, 315, L109

\bibitem{}
Neufeld, D.A., Lepp, S., \& Melnick, G.J. 1995, ApJS, 100, 132

\bibitem{}
Omont, A., Petitjean, P., Guilloteau, S., McMahon, R.G., 
Solomon, P.M., \& Pecontal, E. 1996, Nature, 382, 428

\bibitem{}
Spinoglio, L., \& Malkan, M.A. 1992, ApJ, 399, 504

\bibitem{}
van Dishoeck, E.F., Wright, C., Cernicharo, J., Gonzalez-Alfonso, E.,
de Graauw, T., Helmich, F.P., Vandenbussche, 1998, ApJ, 502, L173

\bibitem{}
Voit, M. 1992, ApJ, 339, 495


\end{references}
\end{document}